\documentclass[preprint]{aastex}

\shorttitle{Forecasting Cloud Cover and Seeing by the GFS model}
\shortauthors{Ye}

\begin{document}


\title{Forecasting Cloud Cover and Atmospheric Seeing for Astronomical Observing: Application and Evaluation of the Global Forecast System}

\author{Q.-z. Ye\altaffilmark{1}}
\affil{Department of Atmospheric Sciences, School of Environmental Science and Engineering, Sun Yat-sen University, Guangzhou, China}

\email{tom6740@gmail.com}

\altaffiltext{1}{Present address: 404, 12 Huasheng St, Guangzhou, 510620 China}

\begin{abstract}
To explore the issue of performing a non-interactive numerical weather forecast with an operational global model in assist of astronomical observing, we use the Xu-Randall cloud scheme and the Trinquet-Vernin AXP seeing model with the global numerical output from the Global Forecast System to generate 3-72h forecasts for cloud coverage and atmospheric seeing, and compare them with sequence observations from 9 sites from different regions of the world with different climatic background in the period of January 2008 to December 2009. The evaluation shows that the proportion of prefect forecast of cloud cover forecast varies from $\sim50$\% to $\sim85$\%. The probability of cloud detection is estimated to be around $\sim30$\% to $\sim90$\%, while the false alarm rate is generally moderate and is much lower than the probability of detection in most cases. The seeing forecast has a moderate mean difference (absolute mean difference $<0.3$" in most cases) and root-mean-square-error or RMSE (0.2"-0.4" in most cases) comparing with the observation. The probability of forecast with $<30$\% error varies between 40\% to 50\% for entire atmosphere forecast and 30\% to 50\% for free atmosphere forecast for almost all sites, which being placed in the better cluster among major seeing models. However, the forecast errors are quite large for a few particular sites.  Further analysis suggests that the error might primarily be caused by the poor capability of GFS/AXP model to simulate the effect of turbulence near ground and on sub-kilometer scale. In all, although the quality of the GFS model forecast may not be comparable with the human-participated forecast at this moment, our study has illustrated its suitability for basic observing reference, and has proposed its potential to gain better performance with additional efforts on model refinement.
\end{abstract}

\keywords{Astronomical Phenomena and Seeing: general --- Astronomical Phenomena and Seeing: individual(astronomical observing, numerical meteorological forecast, cloud cover, astronomical seeing)}

\section{Introduction}

Almost all kinds of ground-based astronomical observations, especially for optical ones, are extremely dependent on meteorological condition, so it is no doubt that making the meteorological prediction as precise as possible would significantly help observers to schedule the observation and improve the efficiency of telescope operation. Among all meteorological variables, cloud amount or cloud cover is apparently the dominant factor, while atmospheric/astronomical seeing, or sometimes described as $Fried$ parameter \citep{fri65}, is also important.

A climatologically study to the annual values of cloud cover and/or atmospheric seeing for a proposed professional observatory is usually done in the ``site survey'' prior to the construction \citep[see][for examples]{wal70,fue90}. However, to maximize the observing resources, astronomers not only need to know the approximate percentage of clear nights in a year, but also wish to know whether the sky will be clear or not in the next a few nights. In another word, they expect weather forecast to be as accurate and precise as possible. However, since astronomical observatories are generally built in distant areas, therefore with sparse meteorological observations available and less interests from meteorologists, special forecast that aim at assisting astronomical observation had not been widely practiced until very recent years. Since the end of 1990s, special forecast services and products based on meso-scale regional numeric models and/or real-time satellite images have been developed at the large professional observatories, such as Mauna Kea Weather Center or MKWC \citep[see][for an overview]{bus02} and the nowcast model at European Southern Observatory (ESO) \citep{era01}.

As the operation of high resolution numeric models would require the ability to perform speedy computation (which is usually only available at large professional observatories), attempts aiming at making direct uses of the model ``fields'' from global/continental models were carried out later, such as the Clear Sky Chart\footnote{\url{http://cleardarksky.com/csk/}} that uses Canadian Meteorological Centre's Global Element Multi-scale (GEM)\footnote{\url{http://collaboration.cmc.ec.gc.ca/science/rpn/gef\_html\_public}} since 2002 \citep{dan03}, and our 7Timer system\footnote{\url{http://7timer.y234.cn}} that uses National Centers for Environmental Prediction (NCEP)'s Global Forecast System\footnote{\url{http://www.emc.ncep.noaa.gov/modelinfo/}} \citep[refer as GFS hereafter; see][for an overview of the model]{sel80,whi08} since 2005. These ``direct-from-model'' forecasts only have decent spatial and vertical resolution (for a comparison, the spatial resolution of the GFS model is about 40km, while the regional model operating at MKWC can reach 1km), but it doesn't require heavy computation works either: the retrieval of the model fields can be done with an Internet-connected Personal Computer (PC) within a couple of minutes, making it the most favorable and probably the only choice when speedy computer is not available and the demand on forecast precision/accuracy is not critical.

Interestingly, although these services have been put into good use by private, public and even some professional observatories, there is no quantitative and systematic understanding on how accurate the model fields are up to now. For example, the only reported estimation of the accuracy of cloud field forecast for a global model was done by \citet{era01} in 1992-1993, which suggested that only 15-25\% of cloudy nights could be identified with the European Centre for Medium-rage Weather Forecasting model (ECMWF)\footnote{\url{http://www.ecmwf.int/}}. This is age old considering there had been a number of significant upgrades of the global models in the following decade. The forecast for atmospheric seeing, on the other hand, is more complex, since it is related to vertical ``fine structure'' of the atmospheric column and is not directly provided as part of output in any global model. In general, there are two tracks for atmospheric seeing forecast: ``nowcast'' track using near real-time meteorological observation profile combining with a statistical model \citep[such as][]{mur95}; or the ``model'' track either using the derivations of the Tatarski's formula \citep{tat61,cou88} or the numeric model proposed by \citet{cou86}. The first track is relatively intuitive and is accurate enough in many occasions, but it has a very short forecast range (usually less than 24h) and heavily depends on the availability and quality of the observational data; for the second track, one would need to divide the atmospheric column into a good number of layers to gain a numeric simulation close enough to the actual situation, which will again require assistance from a speedy computer. In order to solve these shortcomings, \citet{tri06} takes advantages from both tracks and introduced the AXP model. With that model, one only needs to divide the atmospheric column by a number close to that available in most global models, and the consistency from the simulation of the AXP model to the actual situation is satisfying according to the authors. In all, these ``direct-from-model'' forecasts can be a practical solution for the observers without the ability to operate a high precision regional model by their own, and the job to do is to assess how accurate these forecasts are.

We organize this paper as follow. In Section 2, we briefly outline the technical details of the GFS model that used in this study as well as the Xu-Randall cloud scheme that it used for cloud simulation and the AXP model that we used to derive forecast of atmospheric seeing. In Section 3, we describe the observations we used to evaluate the GFS model. Section 4 presents the details and discussions of evaluation methodology and result; while Section 5 gives the concluding remarks of this study.

\section{Forecast}

\subsection{The NCEP GFS model}

The GFS model provides output in two grids with different spatial resolution: grid 003 at $1^{\circ}\times1^{\circ}$, and grid 004 at $0.5^{\circ}\times0.5^{\circ}$. To provide best-possible forecast, we use the later in our study. Model outputs from the period of January 1, 2008 to December 31, 2009 at three-hourly interval for $0<\tau\leq72$h at 00Z initialization is retrieved from the National Operational Model Archive \& Distribution System \citep[NOMADS; see][]{rut06} for evaluation.

The GFS dataset contains approximately 140 fields, supplying forecast fields both of general meteorological interests (such as temperature, humidity, wind direction and speed, etc.) and for special purpose, including cloud cover fraction on different layer (low, mid, high, convective, and of total atmospheric column). Although the atmospheric seeing is not among the output fields, it can be derived indirectly as every of the required meteorological variables are given.

\subsection{Cloud scheme}

In the GFS model, cloud cover fraction for each grid box is computed using the cloud scheme presented by \citet{xu96}, which is shown as eq. [1]. In this equation, $RH$ is the relative humidity, $q^{*}$ and $q_{c}$ are the saturation specific humidity and $q_{cmin}$ is a prescribed minimum threshold value of $q_{c}$. Depending on the environmental temperature, $q^{*}$ and $q_{c}$ are calculated with respect to water phase or ice phase (Yang, personal communication). Cloud cover fraction can therefore be calculated for any layer as long as the $RH$, $q^{*}$ and $q_{c}$ are known and $q_{cmin}$ is suitably prescribed. We note that the calculation is done as part of the model simulation at NCEP, so the cloud fields are used as-is from the GFS datasets.

\begin{equation}
C= \max[RH^{\frac{1}{4}}(1-e^{-\frac{2000(q_{c}-q_{c\min})}{\min\{\max[((1-RH)q^{*})^{\frac{1}{4}},0.0001],1.0\}}}),0.0]
\end{equation}

The GFS model divides the whole atmospheric column into 26 layers. The total cloud cover for entire atmospheric column is derived under the assumption that clouds in all layers are maximally randomly overlapped \citep{yan05}.

\subsection{Seeing model}

The way atmospheric optical turbulence affects astronomical observing is theoretically described by the Kolmogorov-Tatarski turbulence model \citep{tat61,rod81,tok02} which suggested only one parameter is needed to describe the quality of atmospheric seeing $\epsilon_{0}$ with $\lambda$ to be the wavelength associated with seeing ($\lambda=5\times10^{-7}$m in most cases) and $r_{0}$ to be the $Fried$ parameter::

\begin{equation}
\epsilon_{0}=0.98\frac{\lambda}{r_{0}}
\end{equation}

The $Fried$ parameter, $r_{0}$, is defined as followed in the direction of zenith \citep{cou85}, with $Z_{0}$ to be the geopotential height of the observing site, $C_{N}^{2}$ to be the refractive index structure coefficient indicating the strength of turbulence associated with the temperature structure coefficient $C_{T}^{2}$, $P$ to be the pressure in hPa and $T$ to be the temperature in Kelvin:

\begin{equation}
r_{0}=[0.423(\frac{2\pi}{\lambda})^{2}\int_{Z_{0}}^{\infty}C_{N}^{2}dZ]^{-\frac{3}{5}}
\end{equation}
\begin{equation}
C_{N}^{2}=C_{T}^{2}(\frac{7.9\times10^{-5}P}{T^2})^{2}
\end{equation}

Therefore, with the above formulas and suitable inputs, we can derive the total effect of atmospheric turbulence from the integral of $C_{N}^{2}(Z)$ for all atmospheric layers.

\begin{equation}
C_{T}^{2}=\frac{T(\textbf{x})-T(\textbf{x}+\textbf{r})}{|\textbf{r}|^\frac{2}{3}}
\end{equation}

There are several ways to derive $C_{T}^{2}$. A theoretical approach is shown as eq. [5], where \textbf{x} and \textbf{r} to be the position and separation vector, respectively. However, by theory $|\textbf{r}|$ should be around a few 0.1m to precisely describe the effect of turbulence (which is not practically possible to date for model simulation as it would require one to use about 100,000 layers for model simulation). The AXP model using an alternative approach by considering a simple expression of $C_{T}^{2}(h)$ as follow, with the power $p(h)$ adjusts the amplitude of peaks and $A(h)$ connects the level:

\begin{equation}
C_{T}^{2}=\langle C_{T}^{2}\rangle(h)[A(h)\frac{d\bar{\theta}}{dz}]^{p(h)}
\end{equation}

The potential temperature $\theta$ in eq. [6] can be calculated by Poisson's equation with $T$ to be the absolute temperature of a parcel in Kelvin and $P$ to be the pressure of that air parcel in hPa:

\begin{equation}
\theta=T(\frac{1000}{P})^{0.286}
\end{equation}

The authors of the AXP model then statistically determine the values of the three coefficients, $\langle C_{T}^{2}\rangle(h)$, $A(h)$, and $p(h)$, by a vertical spacing of 1km up to an altitude of 30km, based on airborne observations from 162 flights at 9 sites during 1990-2002. On the other hand, the coordinate system adopted by the GFS model is P-coordinate system, which divides the atmospheric column by pressure. The vertical spacing of the GFS model over low and mid level atmosphere is around 1km and is roughly compatible with the Z-coordinate system adopted by the AXP model, however the former becomes too sparse at high level atmosphere (as illustrated in Table~\ref{tbl-1}). To solve this problem, we set up a ``degeneracy'' scheme to allow using the AXP model coefficients in P-coordinate system by weighting the values according to the correlation between atmospheric pressure and altitude defined by U.S. Standard Atmosphere in 1976\footnote{See \url{http://ntrs.nasa.gov/archive/nasa/casi.ntrs.nasa.gov/19770009539\_1977009539.pdf}.}. The transformed coefficients for each pressure layer are given in Table~\ref{tbl-2}. As the output of GFS model provides fields of temperature and atmospheric pressure for each of its vertical layer, we can align the fields with eq. [2], eq. [3], eq. [4], eq. [6] and eq. [7] to derive final $\epsilon_{0}$ in form of arcsec. We refer this hybrid model as GFS/AXP model hereafter.

Trinquet \& Vernin reported an accuracy of 58\% of the original AXP model forecasts with error within $\pm30$\% from observations. However, as we have modified the model layers to fit the GFS model, some additional errors may have been induced. One may expect two possible sources of errors caused by the degeneracy. The first kind of possible error comes from the layers higher than 10km. As the GFS model layers with altitude $>10$km have a vertical thickness significantly larger than 1km, one may suggest that some performance may be lost due to data roughness. However, we argue that the loss of performance by this reason should be minimal, as the high level atmosphere is much less active than the mid or low level atmosphere and contribute minimal turbulence to seeing than the later two\footnote{Study by \citet{li03} suggests the atmosphere beyond 100hPa ($\sim15$,000m) contributes less than 0.01" to the seeing.}, so a vertical spacing degeneration from 1km to 2km at high level atmosphere is unlikely to induce error of significance. On the other hand, the degeneracy in planetary boundary layer or PBL, which serves as the second possible source of error, may induce something large. PBL is a layer that directly influenced by the atmosphere-ground interaction, and it has been shown that the PBL turbulence contributes a major part to the seeing during the night \citep{aba04}. To improve the model performance at PBL, the AXP model divides the planetary boundary layer (PBL) into three sub-layers: 0-50m above ground, 50-100m above ground, and 100-1,000m above ground. However, the GFS model only provides three near-ground fields, which is at 2m above ground, $0.995\sigma$ ($\sim100$m above ground) and 30hPa above ground ($\sim400$m), we have to use identical $d\bar{\theta}/dz$ computed from 2m above ground and $0.995\sigma$ for layers of 0-50m and 50-100m and the $d\bar{\theta}/dz$ computed from $0.995\sigma$ and 30mb above ground for layer of 100-1,000m, the ignorance of data at 50m would almost certainly to induce some error. To work around this issue and allow a direct assess of the GFS/AXP model, we also generate seeing forecast for free atmosphere only - i.e. with PBL ($<100$m in our study) excluded\footnote{We do not exclude the 100m-1,000m region in our free atmosphere seeing forecast because the MASS instrument using for measurement of ``free atmosphere'' seeing considers an altitude of 500m above ground as PBL top limit, and includes measurements of layers at an altitude as low as 500m and 1,000m above ground.} - for our evaluation. The free atmosphere forecast will be compared with the observations from Multi-Aperture Scintillation Sensor (MASS) which is designed to measure the seeing in free atmosphere \citep{tok02}.

\section{Observation}

We collect cloud cover and seeing observations for a total of nine sites from Central/East Asia, Hawaii and Central/South America in the period of January 2008 to December 2009. The respective information of each site including the PBL top limit and $P(h)$ used in the GFS/AXP model are listed in Table~\ref{tbl-3}.

As these observations are all made in sequence with sampling frequency around 1-1.5min except Nanshan and Lulin (which will be deal separately and will be described below), they are firstly processed to match the time interval of the GFS model output (which is 3h). To ensure that the observation is representative in the corresponding interval, we set a minimum data points of 100 for each interval. A sampling frequency of 1-1.5min corresponds 120-180 data points per 3h, so the minimum threshold of 100 is reasonable.

As all seeing observations are obtained either by Differential Image Motion Monitor (DIMM) or MASS, they can be used without further reduction since they give the measurement of $\epsilon_{0}$ directly. On the other hand, the cloud cover observations from Paranal, Nanshan and Lulin are obtained with different tools, so they must be reduced to the same definition with the GFS model output before comparing them with the later. The reduction procedure is described below.

\paragraph{Paranal}

The cloud sensor installed at Paranal determines the sky condition by measuring the flux variation of a star. The sensor graph will suggests a possible ``cloudy'' condition when the root-mean-square (RMS) of the flux variation is larger than 0.02\footnote{See \url{http://archive.eso.org/asm/ambient-server}.}.

\paragraph{Nanshan}

For Nanshan, the operation log is used to verify the sky condition. The operation log includes the image sequence log and observer's notes. We divide each night into two ``parts'': evening and morning. A ``part'' with roughly $>50$\% observable time (with images taken and indication of good observing condition from the observer) would be marked as ``clear'', otherwise it would be marked as ``cloudy''.

\paragraph{Lulin}

A Diffraction Limited Boltwood Cloud Sensor is installed at Lulin to produce sequence observations. The cloud sensor determines the sky condition by comparing the temperature of the sky to ambient ground level temperature, and a difference threshold of $-25^{\circ}$C is set to distinguish ``clear'' and ``cloudy'' conditions\footnote{See \url{http://www.cyanogen.com/downloads/files/claritymanual.pdf}.}.

\section{Evaluation}

\subsection{Evaluation of cloud cover forecast}

The preliminary result from our companion study suggests that roughly 70\% cloud cover forecasts from the GFS model can achieve an error less than 30\% (Ye \& Chen, in prep.) at 0h$<\tau\leq72$h at grid with spacing of $\sim300$km. However, the cloud cover observations used in this study are all made at $single$ $geographic$ $points$ and are $categorical$ (either clear or cloudy) rather than over a grid area and being quantitative. So in order to make comparison between the forecasts and the observations, we first need to simplify the forecast into two categories by setting a spatial threshold that divides ``cloudy'' and ``clear'' situations. To examine the result sensitivity with different thresholds, we set the threshold at 30\%, 50\% and 80\%, respectively. Considering most astronomical observatories are placed at areas with higher chance to be clear rather than cloudy, we should focus the ``cloudy'' event in our study rather than the ``clear'' event. Thus, event ``cloudy'' is set to be the ``yes'' statement, while ``clear'' is set to be the ``no'' statement.

We use four indicators to evaluate the performance of the forecast\footnote{A more detail technical document about the application of these four indicators in meteorology may be found at \url{http://www.ecmwf.int/products/forecasts/guide/Hit\_rate\_and\_False\_alarm\_rate.html}.}: Proportion of Perfect Forecasts (PPF), Probability of Detection (POD), False Alarm Rate (FAR), and Frequency Bias Index (FBI). Let $H$ donates ``hits'' (forecasted and observed), $F$ donates ``false alarms'' (forecasted but not observed), $M$ donates ``missed'' (not forecasted but observed), and $Z$ donates to not-forecasted and not-observed situation, we have:
\begin{displaymath}
PPF=\frac{H+Z}{H+F+M+Z}
\end{displaymath}
\begin{displaymath}
POD=\frac{H}{H+M}
\end{displaymath}
\begin{displaymath}
FAR=\frac{F}{F+Z}
\end{displaymath}
\begin{displaymath}
FBI=\frac{H+F}{H+M}
\end{displaymath}

The $H$, $F$, $M$ and $Z$ numbers for each of the three sites are listed in Table~\ref{tbl-4}, and the comparison results between the forecast and observation are shown in Table~\ref{tbl-5}.

As the result from our companion study has implied that the accuracy declination with the increase of $\tau$ is small (Root-mean-square-error or RMSE varies within 5\% for $\tau$ up to 72h) for the GFS model, we see no need to list percentages for each night separately. The percentages given in Table~\ref{tbl-5} are the average of 0h$<\tau\leq72$h, with the uncertainty ranges donate by the nights with highest/lowest percentages. The large uncertainty of the FBI of Paranal is caused by small dominators since the numbers of ``false alarms'' are 5-10 times more than the numbers of ``misses''.

Although PPF is a biased score since it is strongly influenced by the more common category, it gives a rough indication about the typical percentage of ``correct'' forecast for one site. A clear feature revealed by Table~\ref{tbl-5} is the climate-dependence of the PPF: arid sites tend to have higher PPF, while humid sites tend to have lower. Generally speaking, the PPF varies from ~50\% to ~85\% for the three sites in our sample. As the sites have covered both the extremes of climate type (from arid and humid), it can be considered that this result is relatively representive.

The result revealed by POD is also encouraging. Even for Paranal, the site with an annual cloudy percentage as low as 10\% \citep{ard90}, the POD can still varies between 20-60\% and being better than the 15-25\% percentage reported by \citep{era01}. We also note that the forecast FAR for Paranal and Nanshan is less than half of the forecast POD, far from as good as the human-participated forecasts (for reference, the MKWC forecaster which can reach a FAR as low as 1\%\footnote{See http://mkwc.ifa.hawaii.edu/forecast/mko/stats/index.cgi?night$=$1\&fcster$=$fcsts\&var$=$fog\&cut$=$2.}), but is still reasonable for basic observing reference. We note that the forecast FAR for Lulin is very large; however, considering that Lulin is surrounded by a very complex terrain, where the elevation variation in its belonged grid box is as large as 3,500m, it is not unexpected for a badly-looking FAR. In final, the FBI for all sites are larger than 1, suggesting the GFS model tends to make more ``false alarms'' rather than ``misses''.

\subsection{Evaluation of seeing forecast}

The evaluation of seeing forecast is relatively easier than that of cloud cover forecast since the forecasts and observations are in same definition, and we can simply compare the mean and root-mean-square-error (RMSE) of the difference between the forecasts and observations (Table~\ref{tbl-6}). The mean errors are computed by averaging the differences between instantaneous forecasts and observations. As the increase of error is small ($\Delta$RMSE$<\sim0.05$") for $\tau$ up to 72h, we combine forecasts and observations from all the three nights into a whole for our study.

\begin{figure}
\plotone{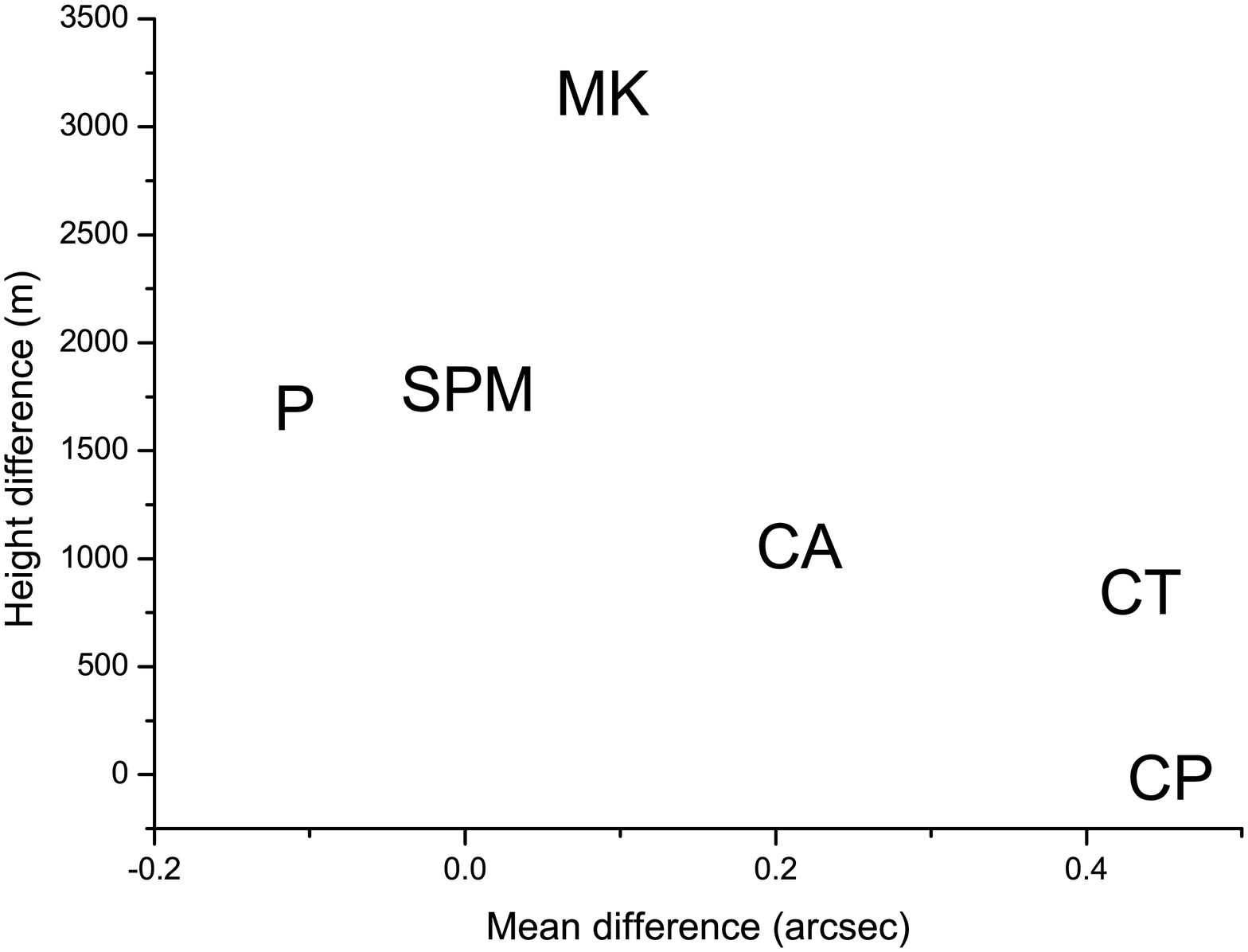}
\caption{Absolute forecast mean errors and the height differences between actual height and GFS grid for all sites for entire atmosphere (upper) and free atmosphere (lower). Abbreviations are: P - Paranal, SPM - San Petro M\'{a}rtir, MK - Mauna Kea, CO - Cerro Tololo, CA - Cerro Armazones, CT - Cerro Tolonchar, and CP - Cerro Pach\'{o}n.\label{fig-1}}
\end{figure}

As seen in Table~\ref{tbl-6}, either entire atmosphere forecast or free atmosphere forecast has moderate error comparing with the observation, while the former tends to slightly overestimate the value and the latter tends to the opposition. However, the tendency is severe for the cases such as Cerro Pach\'{o}n which the mean difference reads +0.46". A possible explanation is that many observatories are located in mountainous areas and being actually much higher than the height of its belonged grid used in GFS models (the height ``differences'' vary from 0-4,000m among our sample, as shown in Table~\ref{tbl-3}); this may produce error in PBL seeing estimation. However, we find no substantial support for this assumption as there is no significantly tendency between the forecast bias and the height difference (Figure~\ref{fig-1}). Dramatically, the site with smallest height difference (Cerro Pach\'{o}n, with a height difference of 9m) has the largest mean error (+0.46").

\begin{figure}
\plotone{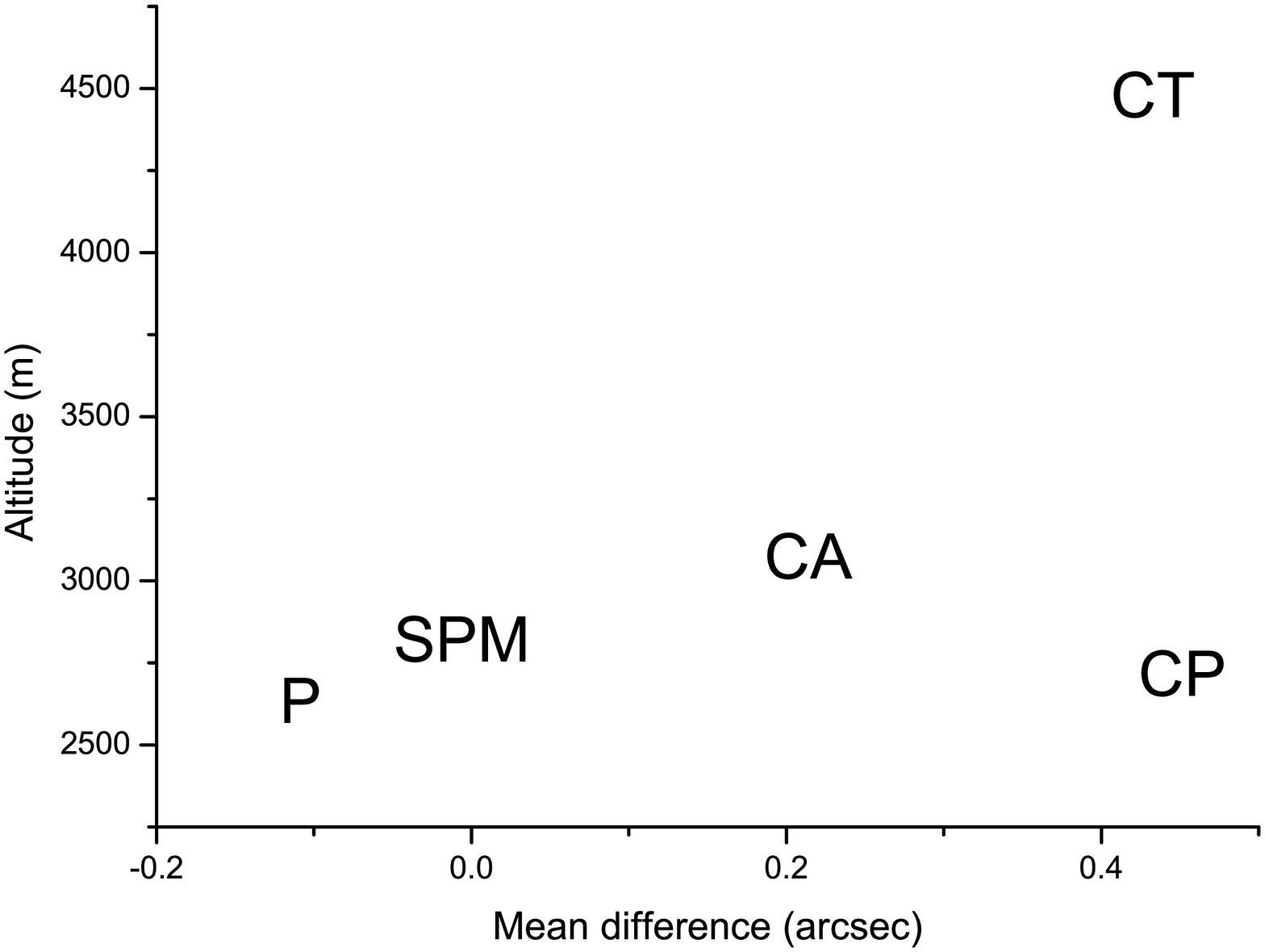}
\caption{Absolute forecast mean error and the altitude of the site for all sites with DIMM observation (upper) and MASS observation (lower). Abbreviations are: P - Paranal, SPM - San Petro M\'{a}rtir, MK - Mauna Kea, CO - Cerro Tololo, CA - Cerro Armazones, CT - Cerro Tolonchar, and CP - Cerro Pach\'{o}n.\label{fig-2}}
\end{figure}

Another equally plausible explanation is the combined effect of the poor consistency between the GFS/AXP model versus the actual $C_{N}^{2}$ variance in PBL and error induced by the layer degeneration of the AXP model for high altitude areas. In fact, we do find a weak bias tendency for free atmosphere forecast as shown in Figure~\ref{fig-2}: the forecast for low-altitude sites tends to be better and all sites below 3,000m have an absolute mean forecast error below 0.25". This feature fits the fact that the airborne data used to determine the coefficients in the AXP model only includes sites with altitude up to 2,835m. But generally speaking, no substantial correlation between bias tendency and an unique geographic modeling factor can be identified, therefore we can only suggest that the bias is contributed by a combination effect of some factors.

\begin{figure}
\plotone{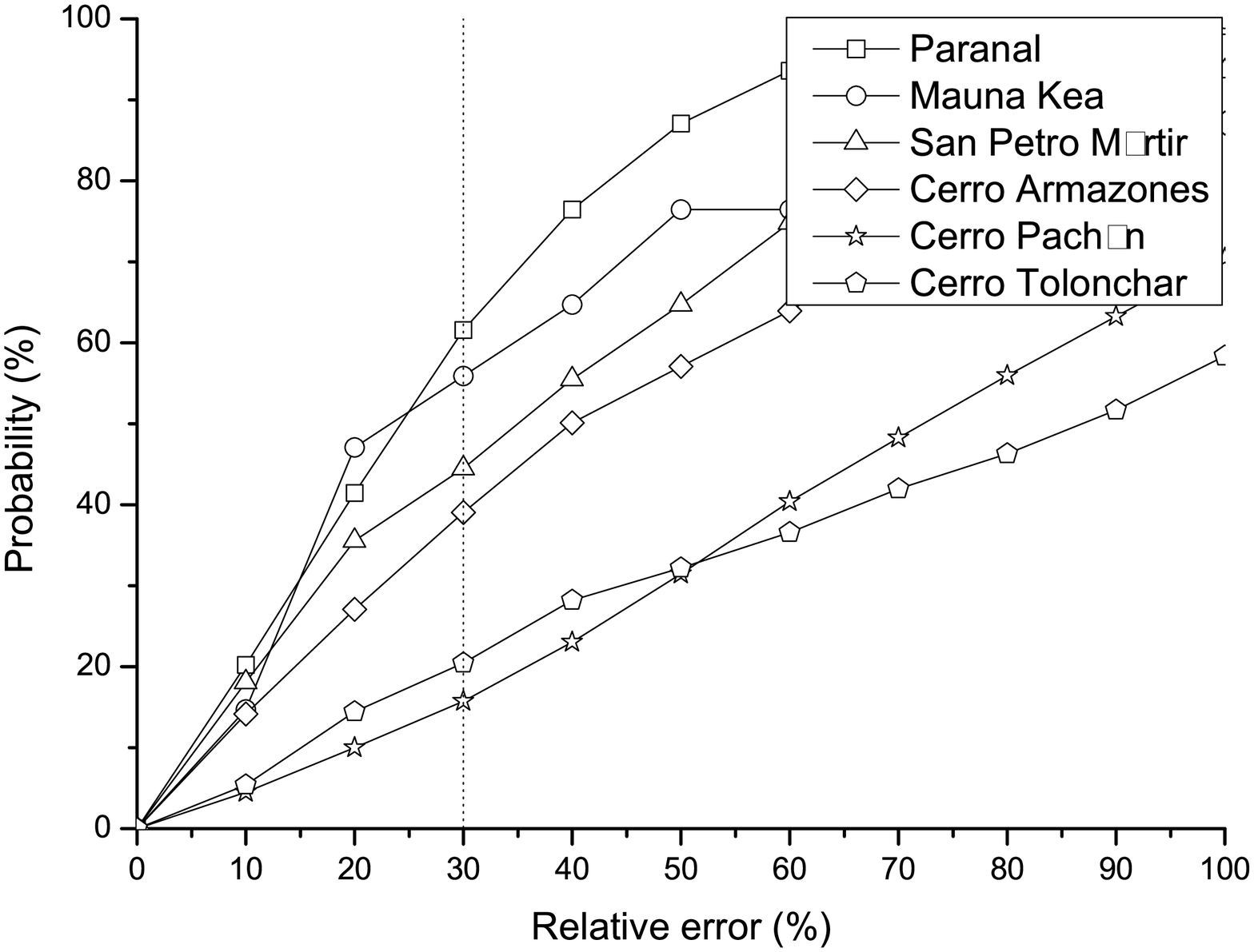}
\caption{Cumulative distribution of relative error of seeing forecast for ground layer plus free atmosphere. Dashed line represents a relative error of 30\%. The probability of $<30$\% forecast error is 63\% for Paranal, 47\% for Mauna Kea, 45\% for San Pedro M\'{a}rtir, 40\% for Cerro Armazones, 14\% for Cerro Pach\'{o}n, and 29\% for Cerro Tolonchar, respectively. The Mauna Kea curve is quite abnormal due to small sample number ($n=34$) and we estimate that the $<30$\% forecast error probability is about 15\% too low.\label{fig-3}}
\end{figure}

\begin{figure}
\plotone{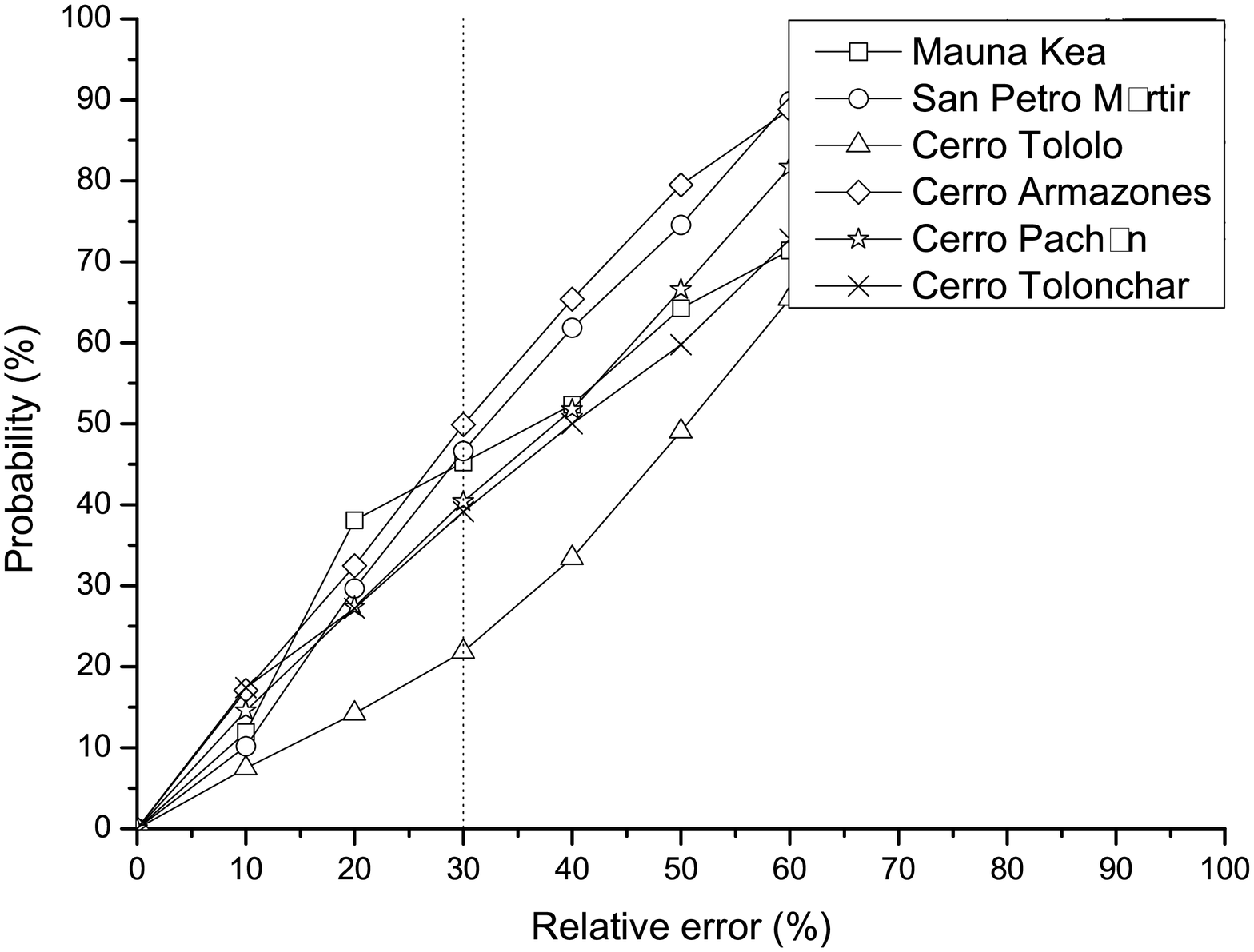}
\caption{Cumulative distribution of relative error of seeing forecast for free atmosphere only. Dashed line represents a relative error of 30\%. The probability of $<30$\% forecast error is 21\% for Mauna Kea, 47\% for San Pedro M\'{a}rtir, 30\% for Cerro Tololo, 38\% for Cerro Armazones, 40\% for Cerro Pach\'{o}n, and 34\% for Cerro Tolonchar, respectively. The Mauna Kea curve is quite abnormal due to small sample number ($n=42$) and we estimate that the $<30$\% forecast error probability is about 15\% too low.\label{fig-4}}
\end{figure}

To give a more comprehensive understanding on the quality of the GFS/AXP forecast, we plot the cumulative distribution of the relative forecast error for each site as in Figure~\ref{fig-3} (forecast for entire atmosphere) and Figure~\ref{fig-4} (forecast for free atmosphere only). Let $\hat{\epsilon_{0}}$ to be the forecasted seeing value and $\epsilon_{0}$ to be the observed seeing value, the relative forecast error $E(\epsilon_{0})$ is calculated by

\begin{displaymath}
E(\epsilon_{0})=\frac{\mid\hat{\epsilon_{0}}-\epsilon_{0}\mid}{\epsilon_{0}}
\end{displaymath}

From the two figures, we can identify that the probabilities of producing forecast with $<30$\% error concentrate in 40-50\% for entire atmosphere seeing forecast and 30-50\% for free atmosphere seeing forecast. By contrast, Trinquet \& Vernin gives a probability of 58\% and 50\% for the original AXP model to produce entire atmosphere or free atmosphere forecast in the same quality. Generally speaking, our result is in rough agreement with them, although the model's performance is rather unsatisfying on a few particular sites.

\begin{figure}
\plotone{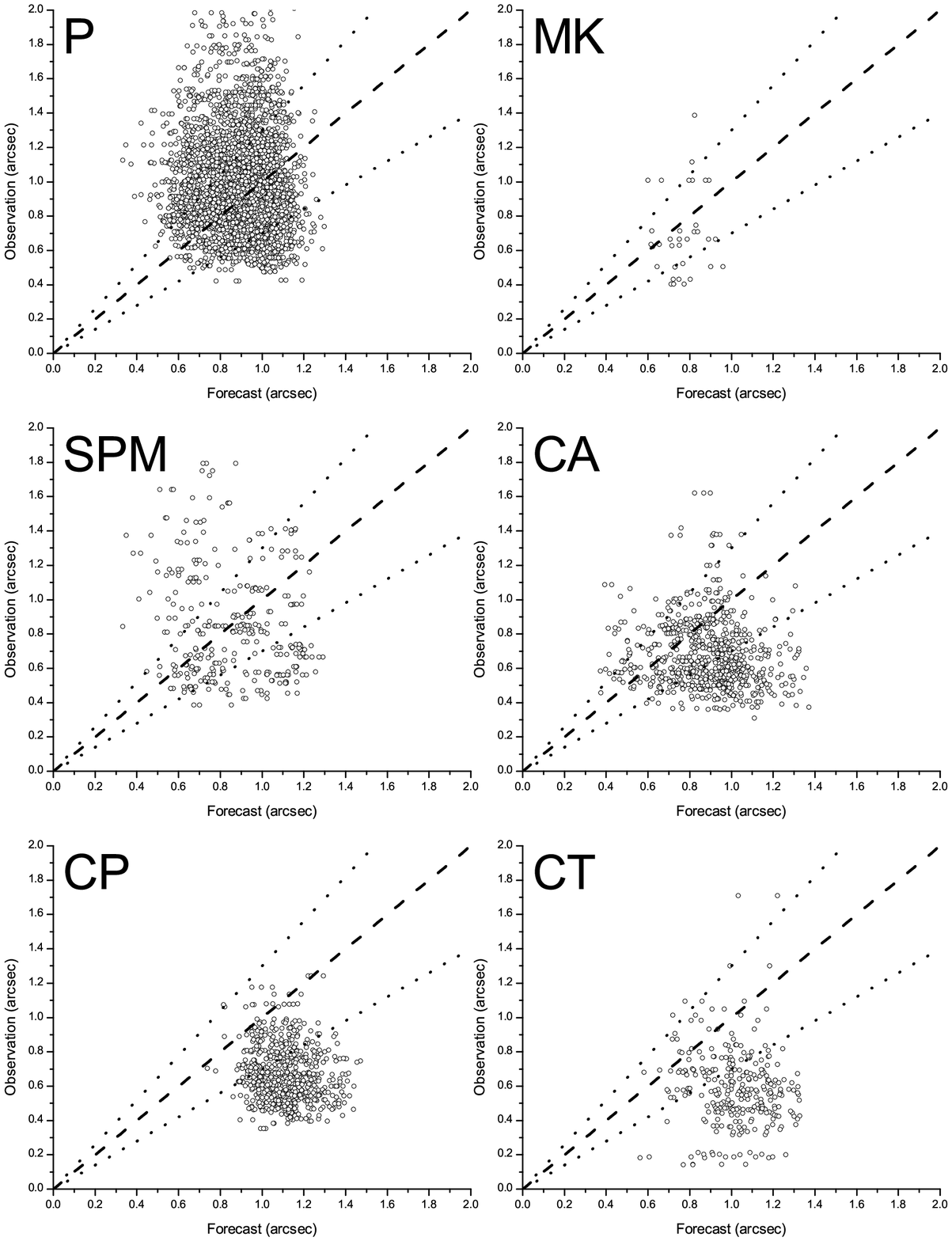}
\caption{Forecast-by-observation figure of entire atmosphere seeing forecast for Paranal (P), Mauna Kea (MK), San Pedro M\'{a}rtir (SPM), Cerro Armazones (CA), Cerro Pach\'{o}n (CP), and Cerro Tolonchar (CT). Dashed line corresponds to the ideal case (slope=1) while dotted lines are 30\% error uncertainty from the ideal case.\label{fig-5}}
\end{figure}

\begin{figure}
\plotone{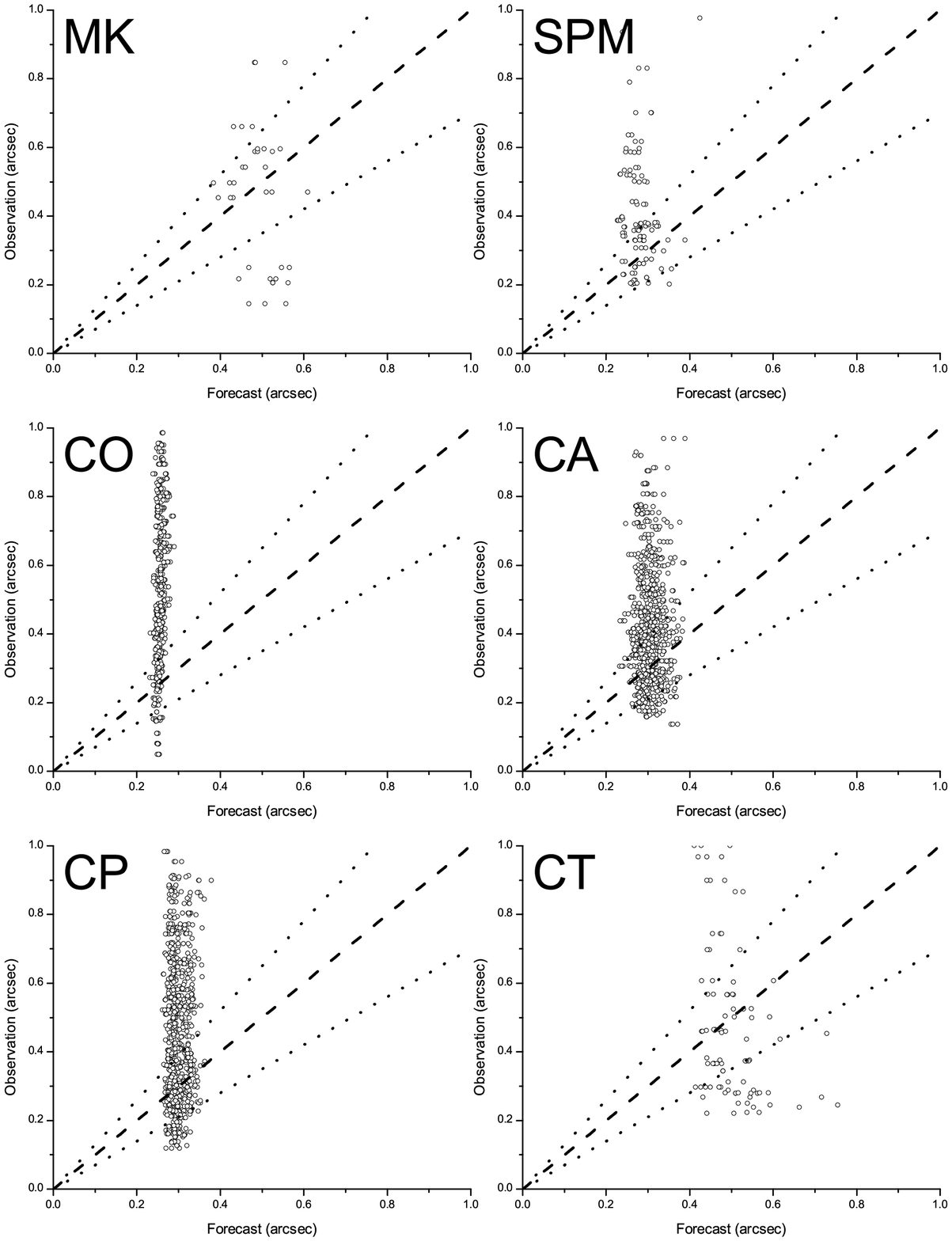}
\caption{Forecast-by-observation figure of free atmosphere seeing forecast for Mauna Kea (MK), San Pedro M\'{a}rtir (SPM), Cerro Tololo (CO), Cerro Armazones (CA), Cerro Pach\'{o}n (CP), and Cerro Tolonchar (CT). Dashed line corresponds to the ideal case (slope=1) while dotted lines are 30\% error uncertainty from the ideal case.\label{fig-6}}
\end{figure}

In addition to the summarizing table and cumulative distribution figures, we also present the forecast-observation distributions for each site as in Figure~\ref{fig-5} and Figure~\ref{fig-6}. We can see that although the statistical values might be satisfying, the forecast-observation distribution figures imply that the correlation between forecast and observation is poor. This is not out-of-expected, since the study of \citet{che09} at MKWC has suggested that one may achieve a good approximation of the actual condition with $\sim80$ vertical layers up to 10hPa, the $\sim15$ layers used in our model may be far from sufficient for a well-correlated distribution. Our assumption is reinforce by the fact that forecasts for free atmosphere seeing are generally concentrated in a narrow region despite their mean is close to that of observation, as shown in Figure~\ref{fig-6}. This phenomenon implies that turbulence over sub-kilometer scale in the vertical direction might be the major contributor to a bad seeing condition in free atmosphere region.

In final, we compare our result with several major models/forecasters (Table~\ref{tbl-7}). The RMSE and $<30$\% error probabilities for the original AXP model, Vernin-Tatarski model, $C_{N}^{2}$/seeing median and seeing mean are derived by Trinquet \& Vernin with their experimental profile observations. The MKWC/WRF model is initiated with the GFS model output, but the simulation would eventually derives output with final grid spacing of 1km and a vertical layer number of 40, resulting a forecast RMSE at 0.36" for the 1st night. As revealed by Table~\ref{tbl-7}, the RMSE uncertainty ranges of GFS/AXP model just fall between the mean of MKWC forecaster and the original AXP model, while the $<30$\% error probability is in the better cluster in all models under most cases. Interestingly, a ``direct'' comparison of the GFS/AXP model with the MKWC forecaster over the seeing forecast for the same site (Mauna Kea) even slightly favors the former (0.26" versus 0.28" for three-night mean). In short, this result has highlighted the potential of the GFS/AXP model to be a competitive forecast tool once the layer degeneracy and high-altitude issues are solved.

\section{Conclusion}

We have carried out a comprehensive study over the topic of performing automatic numeric forecast of cloud cover and atmospheric seeing with the Global Forecast System, an operational global model. Sequence observations on cloud cover and atmospheric seeing from 9 sites from different regions of the world with different climatic background in the period of January 2008 to December 2009 are used to evaluate the forecast. Although the performance of the model forecast may not be comparable with the human-participated forecast, our study has shown that the forecast to be acceptable for basic observing reference. Our study has also reveal the possibility to gain better performance from the model with additional efforts on model refinement.

For cloud cover forecast, we have found that the proportion of perfect forecast varies from $\sim50$\% to $\sim85$\% for all three sites we evaluated, including a site located in subtropical region with a very humid climate (Lulin). In particular, we have found that the model is capable to detect a significant amount of occurring clouds while the false alarm rate is moderate. The probability of detection is still measured to be 20-60\% even for site with very low cloudy probability (Paranal).

For atmospheric seeing forecast, we adopted the AXP model introduced by Trinquet \& Vernin. We found that forecast for entire atmosphere tends to slightly overestimate the seeing, while the free atmosphere forecast tends to the opposition. The RMSE for free atmosphere seeing is smaller (0.22"-0.42") than that of entire atmosphere (0.26"-0.50"), but both values can indicate a decent quality of the forecast comparing with the other major models. Further analysis suggests that a major contributor to the forecast error might be the layer degeneracy issue of the hybrid GFS/AXP model. On the other hand, the probability of GFS/AXP forecasts with $<30$\% error varies between 40-50\% for entire atmosphere forecast and 30-50\% for free atmosphere forecast in most cases, which is in the better cluster among major seeing models. To conclude, our study has suggested a decent performance of the GFS/AXP model that is suitable for basic observing reference. Our study has also suggested that the model has the potential to become a rather useful forecast tool with additional efforts on model refinement.

\acknowledgments

The author wish to thank all people who have given help to this work, particularly to an anonymous reviewer for his/her very constructive comments that led to an advance of this work, and to Sisi Chen, Shaojia Fan, Linjiong Zhou, and Johnson Lau for their helps from various aspects. The author is very grateful to Chenzhou Cui on behalf of the Large Sky Area Multi-Object Fibre Spectroscopic Telescope (LAMOST) team at National Astronomical Observatory, Chinese Academy of Sciences, who made this work possible by providing excellent long-term resources of server and computing for our 7Timer system since 2005. The author also wish to thank Mt. Nanshan Xingming Observatory, National Central University Lulin Observatory, Cerro Tololo Inter-American Observatory, European Southern Observatory, and the Thirty Meter Telescope (TMT) Site Testing Project (Sch\"{o}ck et al, 2009) for providing and/or granting permission to access and use the astro-meteorological observation data from Mt. Nanshan, Lulin, Cerro Pachon, Paranal, and the TMT test sites. In addition, the author would like to thank Xing Gao (No. 1 Senior High School of \"{U}r\"{u}mqi/Xingming Observatory), Hung-Chin Lin (National Central University/Lulin Observatory), Andrei Tokovinin (Cerro Tololo Inter-American Observatory), Edison Bustos (Cerro Tololo Inter-American Observatory), and Felipe Daruich (Gemini Observatory), for providing assistances on obtaining the observational data.

\clearpage

\clearpage




\clearpage

\begin{deluxetable}{cc}
\tablecaption{Layers of the GFS model (in P-coordinate system) and corresponded altitude layers (in Z-coordinate system).\label{tbl-1}}
\tablewidth{0pt}
\tablehead{
\colhead{GFS layer} & \colhead{Corresponding altitude}
}
\startdata
2m above ground&2m above ground\\
$0.995\sigma$\tablenotemark{a}&$\sim100$m above ground\\
30hPa above ground&$\sim400$m above ground\\
900-850hPa&988-1,457m\\
850-800hPa&1,457-1,948m\\
800-750hPa&1,948-2,465m\\
750-700hPa&2,465-3,011m\\
700-650hPa&3,011-3,589m\\
650-600hPa&3,589-4,205m\\
600-550hPa&4,205-4,863m\\
550-500hPa&4,863-5,572m\\
500-450hPa&5,572-6,341m\\
450-400hPa&6,341-7,182m\\
400-350hPa&7,182-8,114m\\
350-300hPa&8,114-9,160m\\
300-250hPa&9,160-10,359m\\
250-200hPa&10,359-11,770m\\
200-150hPa&11,770-13,503m\\
150-100hPa&13,503-15,791m\\
100-70hPa&15,791-17,662m\\
70-50hPa&17,662-19,314m\\
50-30hPa&19,314-21,629m\\
30-20hPa&21,629-23,313m\\
20-10hPa&23,313-25,908m\\
\enddata
\tablenotetext{a}{$0.995\sigma$ stands for the level at $0.995\sigma$ in sigma coordinate system. The sigma coordinate system is a vertical coordinate system being widely used in numerical models. It defines the vertical position as a ratio of the pressure difference between that position and the top of its belonging grid.}
\end{deluxetable}

\begin{deluxetable}{cccccc}
\rotate
\tablecaption{Coefficients for the AXP model as well as the corresponded GFS model layer and weight.\label{tbl-2}}
\tablewidth{0pt}
\tablehead{
\colhead{AXP layer} & \colhead{Corresponding GFS layer and weight} & \colhead{\={P}} & \colhead{$p(h)$} & \colhead{$A(h)$} & \colhead{$\langle C_{T}^{2}\rangle(h)$}
}
\startdata
0-50m above ground & 2m above ground to $0.995\sigma$ & $P(h)$-3hPa & 0.5 & 1.6E+2 & 1.4E-2\\
50-100m above ground & & $P(h)$-9hPa & & & 4.8E-4$(h/100)^{-2.6}$\\
100-1,000m above ground & $0.995\sigma$ to 30hPa above ground & $P(h)$-60hPa & & & 3.4E-5$(h/1000)^{-1.1}$\\
1-2km & 900-850hPa (50\%) & 845hPa & 0.3 & 1.8E+3 & 5.2E-5\\
& 850-800hPa (40\%) & & & & \\
& 800-750hPa (10\%) & & & & \\
2-3km & 800-750hPa (50\%) & 745hPa & 1.3 & 5.6E+2 & 4.2E-5 \\
& 750-700hPa (50\%) & & & & \\
3-4km & 700-650hPa (60\%) & 655hPa & 1.7 & 4.6E+2 & 2.8E-5 \\
& 650-600hPa (40\%) & & & & \\
4-5km & 650-600hPa (20\%) & 575hPa & 1.7 & 3.8E+2 & 2.5E-5 \\
& 600-550hPa (70\%) & & & & \\
& 550-500hPa (10\%) & & & & \\
5-6km & 550-500hPa (60\%) & 505hPa & 2.6 & 2.6E+2 & 1.8E-5 \\
& 500-450hPa (40\%) & & & & \\
6-7km & 500-450hPa (30\%) & 440hPa & 1.1 & 4.6E+2 & 1.5E-5 \\
& 450-400hPa (70\%) & & & & \\
7-8km & 450-400hPa (20\%) & 380hPa & 0.8 & 6.8E+2 & 1.6E-5 \\
& 400-350hPa (80\%) & & & & \\
8-9km & 400-350hPa (10\%) & 330hPa & 0.6 & 1.0E+3 & 1.6E-5 \\
& 350-300hPa (90\%) & & & & \\
9-10km & 350-300hPa (20\%) & 285hPa & 0.3 & 2.2E+3 & 1.9E-5 \\
& 300-250hPa (80\%) & & & & \\
10-11km & 300-250hPa (40\%) & 245hPa & 0.5 & 6.8E+2 & 2.6E-5 \\
& 250-200hPa (60\%) & & & & \\
11-12km & 250-200hPa (80\%) & 210hPa & 0.7 & 3.2E+2 & 3.4E-5 \\
& 200-150hPa (20\%) & & & & \\
12-13km & 200-150hPa (100\%) & 177hPa & 0.6 & 2.2E+2 & 4.4E-5 \\
13-14km & 200-150hPa (50\%) & 150hPa & 0.1 & 8.3E+3 & 4.8E-5 \\
& 150-100hPa (50\%) & & & & \\
14-15km & 150-100hPa (100\%) & 126hPa & 0.2 & 1.0E+3 & 5.5E-5 \\
15-16km & 150-100hPa (80\%) & 105hPa & -0.4 & 3.2E+1 & 6.5E-5 \\
& 100-70hPa (20\%) & & & & \\
16-17km & 100-70hPa (100\%) & 88hPa & -0.3 & 1.5E+1 & 8.3E-5 \\
17-18km & 100-70hPa (70\%) & 72hPa & 2.5 & 5.6E+1 & 1.1E-4 \\
& 70-50hPa (30\%) & & & & \\
18-19km & 70-50hPa (100\%) & 59hPa & -0.9 & 2.6E+1 & 1.1E-4 \\
19-20km & 70-50hPa (30\%) & 48hPa & 3.3 & 3.8E+1 & 9.5E-5 \\
& 50-30hPa (70\%) & & & & \\
20-21km & 50-30hPa (100\%) & 39hPa & -1.0 & 2.6E+1 & 8.2E-5 \\
21-22km & 50-30hPa (60\%) & 31hPa & 1.5 & 4.6E+1 & 7.4E-5 \\
& 30-20hPa (40\%) & & & & \\
22-23km & 30-20hPa (100\%) & 24hPa & -1.9 & 3.2E+1 & 7.5E-5 \\
23-24km & 30-20hPa (30\%) & 19hPa & 1.3 & 4.6E+1 & 8.7E-5 \\
& 20-10hPa (70\%) & & & & \\
24-25km & 20-10hPa (100\%) & 15hPa & 1.1 & 3.8E+1 & 1.1E-4 \\
25-26km & 20-10hPa (100\%) & 11hPa & 1.5 & 3.8E+1 & 1.3E-4 \\
\enddata
\end{deluxetable}

\begin{deluxetable}{ccccccccc}
\rotate
\tabletypesize{\scriptsize}
\tablecaption{Observing sites. PBL top stands for the upper limit of planetary boundary layer (PBL) adopted in the GFS/AXP model. $P(h)$ is the atmospheric pressure at the height of the site used in GFS/AXP model, calculated under standard atmosphere.\label{tbl-3}}
\tablewidth{0pt}
\tablehead{
\colhead{Site} & \colhead{Location} & \colhead{Elevation ($h$)} & \colhead{GFS grid elevation} & \colhead{Climate} & \colhead{Data type} & \colhead{Availability} & \colhead{PBL top} & \colhead{$P(h)$}
}
\startdata
Paranal & -70.40, -24.63 & 2,635m & 919m & Arid & Cloud, DIMM & 2008 Jan. 1 - 2009 Dec. 31 & 3,700m & 734hPa \\
Mauna Kea & -155.48, +19.83 & 4,050m & 896m & Highlands & DIMM, MASS & 2008 Jan. 1- 2008 May 31 & 5,100m & 612hPa \\
San Pedro M\'{a}rtir & -115.47, +31.05 & 2,830m & 1,038m & Arid & DIMM, MASS & 2008 Jan. 1 - 2008 Aug. 31 & 3,800m & 716hPa \\
Cerro Tololo & -70.80, -30.17 & 2,200m & 926m & Arid & MASS & 2008 Jan. 1 - 2009 Dec. 31 & 3,200m & 775hPa \\
Nanshan & +87.18, +43.47 & 2,080m & 2,233m & Semiarid & Cloud & 2008 Jan. 1 - 2009 Jul. 4 & - & - \\
Lulin & +120.87, +23.47 & 2,862m & 1,345m & Humid subtropical & Cloud & 2008 Jan. 1 - 2009 Jul. 31 & - & - \\
Cerro Armazones & -70.20, -24.60 & 3,064m & 1,997m & Arid & DIMM, MASS & 2008 Jan. 1 - 2009 Sep. 30 & 4,100m & 695hPa \\
Cerro Pach\'{o}n & -70.73, -30.23 & 2,715m & 2,706m & Arid & DIMM, MASS & 2008 Jan. 1 - 2009 Dec. 31\tablenotemark{a} & 3,700m & 727hPa \\
Cerro Tolonchar & -67.98, -23.93 & 4,480m & 3,626m & Highlands & DIMM, MASS & 2008 Jan. 1 - 2008 Sep. 30 & 5,500m & 579hPa \\
\enddata
\tablenotetext{a}{Ground-based seeing observation from 2008 Oct. 31 to 2009 Apr. 29 only.}
\end{deluxetable}

\begin{table}
\begin{center}
\caption{Cloud cover forecast result for Paranal, Nanshan, and Lulin.\label{tbl-4}}
\begin{tabular}{ccccc}
\tableline\tableline
Threshold & Category & Paranal & Nanshan & Lulin \\
\tableline
30\% & $H$ & 73 & 715 & 682 \\
& $F$ & 443 & 482 & 860 \\
& $M$ & 71 & 38 & 14 \\
& $Z$ & 1411 & 241 & 94 \\
& Total $n$ & 1998 & 1476 & 1650 \\
\tableline
50\% & $H$ & 62 & 684 & 670 \\
& $F$ & 353 & 386 & 813 \\
& $M$ & 82 & 69 & 26 \\
& $Z$ & 1501 & 337 & 141 \\
& Total $n$ & 1998 & 1476 & 1650 \\
\tableline
80\% & $H$ & 44 & 576 & 642 \\
& $F$ & 225 & 232 & 734 \\
& $M$ & 100 & 177 & 54 \\
& $Z$ & 1629 & 491 & 220 \\
& Total $n$ & 1998 & 1476 & 1650 \\
\tableline
\end{tabular}
\end{center}
\end{table}

\begin{table}
\begin{center}
\caption{Evaluation result of cloud cover forecast for Paranal, Nanshan, and Lulin.\label{tbl-5}}
\begin{tabular}{ccccc}
\tableline\tableline
Threshold & Indicator & Paranal ($n$=1,998) & Nanshan ($n$=1,476) & Lulin ($n$=1,650) \\
\tableline
30\% & PPF & $0.74^{+0.03}_{-0.02}$ & $0.65^{+0.02}_{-0.02}$ & $0.47^{+0.01}_{-0.01}$ \\
& POD & $0.51^{+0.12}_{-0.07}$ & $0.95^{+0.04}_{-0.05}$ & $0.98^{+0.01}_{-0.01}$ \\
& FAR & $0.24^{+0.04}_{-0.03}$ & $0.67^{+0.01}_{-0.01}$ & $0.90^{+0.02}_{-0.01}$ \\
& FBI & $3.58^{+0.61}_{-0.50}$ & $1.59^{+0.03}_{-0.05}$ & $2.22^{+0.01}_{-0.03}$ \\
\tableline
50\% & PPF & $0.78^{+0.03}_{-0.02}$ & $0.69^{+0.03}_{-0.03}$ & $0.49^{+0.02}_{-0.02}$ \\
& POD & $0.43^{+0.07}_{-0.04}$ & $0.91^{+0.05}_{-0.06}$ & $0.96^{+0.03}_{-0.02}$ \\
& FAR & $0.19^{+0.03}_{-0.03}$ & $0.53^{+0.02}_{-0.01}$ & $0.85^{+0.02}_{-0.02}$ \\
& FBI & $2.88^{+0.52}_{-0.48}$ & $1.32^{+0.04}_{-0.04}$ & $2.13^{+0.01}_{-0.02}$ \\
\tableline
80\% & PPF & $0.84^{+0.02}_{-0.03}$ & $0.72^{+0.04}_{-0.03}$ & $0.52^{+0.02}_{-0.02}$ \\
& POD & $0.31^{+0.15}_{-0.10}$ & $0.76^{+0.06}_{-0.06}$ & $0.92^{+0.05}_{-0.02}$ \\
& FAR & $0.12^{+0.05}_{-0.03}$ & $0.32^{+0.01}_{-0.01}$ & $0.77^{+0.02}_{-0.02}$ \\
& FBI & $1.86^{+0.71}_{-0.44}$ & $1.07^{+0.06}_{-0.06}$ & $1.98^{+0.05}_{-0.05}$ \\
\tableline
\end{tabular}
\end{center}
\end{table}

\begin{table}
\begin{center}
\caption{Mean difference (calculated by the mean of forecast minus the mean of observation) and RMSE of seeing forecast for the sample sites.\label{tbl-6}}
\begin{tabular}{ccccccc}
\tableline\tableline
Site & Entire atmosphere & & & Free atmosphere & & \\
& Mean difference & RMSE & Sample $n$ & Mean difference & RMSE & Sample $n$ \\
\tableline
Paranal & -0.09" & 0.36" & 3,630 & & & \\
Mauna Kea & -0.15" & 0.26" & 34 & -0.32" & 0.42" & 42 \\
San Pedro M\'{a}rtir & +0.01" & 0.45" & 326 & -0.10" & 0.22" & 118 \\
Cerro Tololo & & & & -0.24" & 0.34" & 640 \\
Cerro Armazones & +0.20" & 0.34" & 776 & -0.18" & 0.27" & 878 \\
Cerro Pach\'{o}n & +0.46" & 0.50" & 738 & -0.14" & 0.27" & 928 \\
Cerro Tolonchar & +0.29" & 0.40" & 298 & -0.26" & 0.33" & 92 \\
\tableline
\end{tabular}
\end{center}
\end{table}

\begin{table}
\begin{center}
\caption{Comparison between the GFS/AXP model versus several other major seeing models. The values of the models of original AXP, Vernin-Tatarski, $C_{N}^{2}$ median, seeing median, and seeing mean are taken from the work of Trinquet \& Vernin. The probabilities of forecast with $<30$\% error for the cases of MKWC forecaster/WRF model are estimated manually from the plots generated on the MKWC website.\label{tbl-7}}
\begin{tabular}{ccccccc}
\tableline\tableline
Model & Forecast lead time & RMSE & $<30$\% error probability \\
\tableline
GFS/AXP & 1-3 nights & 0.26"-0.50" & Mostly 40-50\% \\
Original AXP & Simultaneous & 0.48" & $\sim58$\% \\
Vernin-Tatarski & Simultaneous & & 30\% \\
$C_{N}^{2}$ median & Simultaneous & & 41\% \\
Seeing median & Simultaneous & & 45\% \\
Seeing mean & Simultaneous & & 41\% \\
MKWC Forecaster\tablenotemark{a} & 1-3 nights & 0.25"-0.30" & $\sim50$\%? \\
MKWC WRF\tablenotemark{b} & 1 night & 0.36" & $\sim50$\%? \\
\tableline
\end{tabular}
\tablenotetext{a}{See http://mkwc.ifa.hawaii.edu/forecast/mko/stats/index.cgi?night$=$1\&fcster$=$fcsts\&var$=$seeing\&cut$=$2.}
\tablenotetext{b}{See http://mkwc.ifa.hawaii.edu/forecast/mko/stats/index.cgi?night$=$1\&fcster$=$wrf\&var$=$seeing\&cut$=$2.}
\end{center}
\end{table}

\end{document}